\documentclass[pre,aps,superscriptaddress,twocolumn,floatfix]{revtex4-1}

\usepackage{dsfont}
\usepackage{mathtext}

\usepackage{mathtools}
\usepackage[usenames,dvipsnames]{xcolor}
\usepackage{amsmath}
\usepackage{amssymb,mathrsfs}
\usepackage{graphicx}
\usepackage{epstopdf}

\usepackage{mathtext}

\usepackage{bm}

\newcommand{\bi}{{\bm i}}
\newcommand{\bj}{{\bm j}}
\newcommand{\br}{{\bm r}}

\newcommand{\tpsi}{\tilde{\psi}}

\begin{document}

\title{Bragg solitons in topological Floquet insulators}

\author{Sergey~K.~Ivanov}
\affiliation{Moscow Institute of Physics and Technology, Institutsky lane 9, Dolgoprudny, Moscow region, 141700, Russia}
\affiliation{Institute of Spectroscopy, Russian Academy of Sciences, Troitsk, Moscow, 108840, Russia}

\author{Yaroslav~V.~Kartashov}
\affiliation{Institute of Spectroscopy, Russian Academy of Sciences, Troitsk, Moscow, 108840, Russia}
\affiliation{ICFO-Institut de Ciencies Fotoniques, The Barcelona Institute of Science and Technology, 08860 Castelldefels (Barcelona), Spain}

\author{Lukas~J.~Maczewsky}
\affiliation{Institute for Physics, University of Rostock, Albert-Einstein-Str. 23, 18059 Rostock, Germany}

\author{Alexander~Szameit}
\affiliation{Institute for Physics, University of Rostock, Albert-Einstein-Str. 23, 18059 Rostock, Germany}

\author{Vladimir~V.~Konotop}
\affiliation{Departamento de F\'isica, Faculdade de Ci\^encias, Universidade de Lisboa, Campo Grande, Ed. C8, Lisboa 1749-016, Portugal}
\affiliation{Centro de F\'isica Te\'orica e Computacional, Universidade de Lisboa, Campo Grande, Ed. C8, Lisboa 1749-016, Portugal}

\begin{abstract}
	We consider a topological Floquet insulator consisting of two honeycomb arrays of identical waveguides having opposite helicities. The interface between the arrays supports two distinct topological edge states, which can be resonantly coupled by additional weak longitudinal refractive index modulation with a period larger than the helix period. 
	In the presence of Kerr nonlinearity such coupled edge states enable topological Bragg solitons. Theory and examples of such solitons are presented.
\end{abstract}

\maketitle
The physics of topological insulators, initially introduced in linear electronic systems (see ~\cite{electron1} for a review), nowadays rapidly expands into the plane of nonlinear materials. 
Many optical \cite{topphot1} and optoelectronic \cite{topphot2} systems, hosting topological effects
attract particular attention. It has been shown that nonlinear effects in such systems
can be used to control excitation and direction of the topological currents \cite{control1,control2,control3,control4}, are crucial for stable operation of topological lasers \cite{laser1,laser2}, give rise to modulational instabilities of the edge states \cite{modinst1,modinst2}, and may lead to the formation of topological quasi-solitons in various optical \cite{optsol1,modinst1,optsol2,optsol3,optsol4} and optoelectronic \cite{polsol1,polsol2} settings. Such quasi-solitons inherit topological protection and localization across the edge of the insulator from linear edge states and remain localized also along the edge, being fully two-dimensional hybrid nonlinear states.

Generally, topological quasi-solitons are introduced as envelope solitons for carrier wave given by linear topological edge state \cite{optsol2,optsol4,polsol1}, and are described by the effective nonlinear Schr\"odinger equation (NLS) with second-order dispersion dictated by the dispersion of corresponding linear edge state giving rise to soliton. However, this is not the only mechanism that can lead to formation of topological quasi-solitons. Rare examples of self-sustained topological quasi-solitons of different physical origin are provided by Bragg solitons in spin-orbit-coupled Bose-Einstein condensates \cite{bragg1} and discrete optical Dirac solitons \cite{dirac1}.

In this Letter, we predict that a new type of topological quasi-soliton - Bragg soliton - can form in Floquet topological insulators. We use a photonic platform based on arrays of helical waveguides that was successfully employed for demonstration of linear Floquet \cite{linfloq1} and anomalous Floquet \cite{linfloq2,linfloq3} insulators.  {Helical arrays can be also induced optically as reported in \cite{optind01}.} Bragg solitons are obtained at the interface of two arrays with waveguides having opposite rotation directions and supporting two different Floquet edge states. When such states are resonantly coupled by additional slow longitudinal modulation, enabling Rabi oscillations of the energy between them, Bragg solitons form in the presence of cubic nonlinearity. 

\begin{figure}[t!]
\centering
\includegraphics[width=1\linewidth]{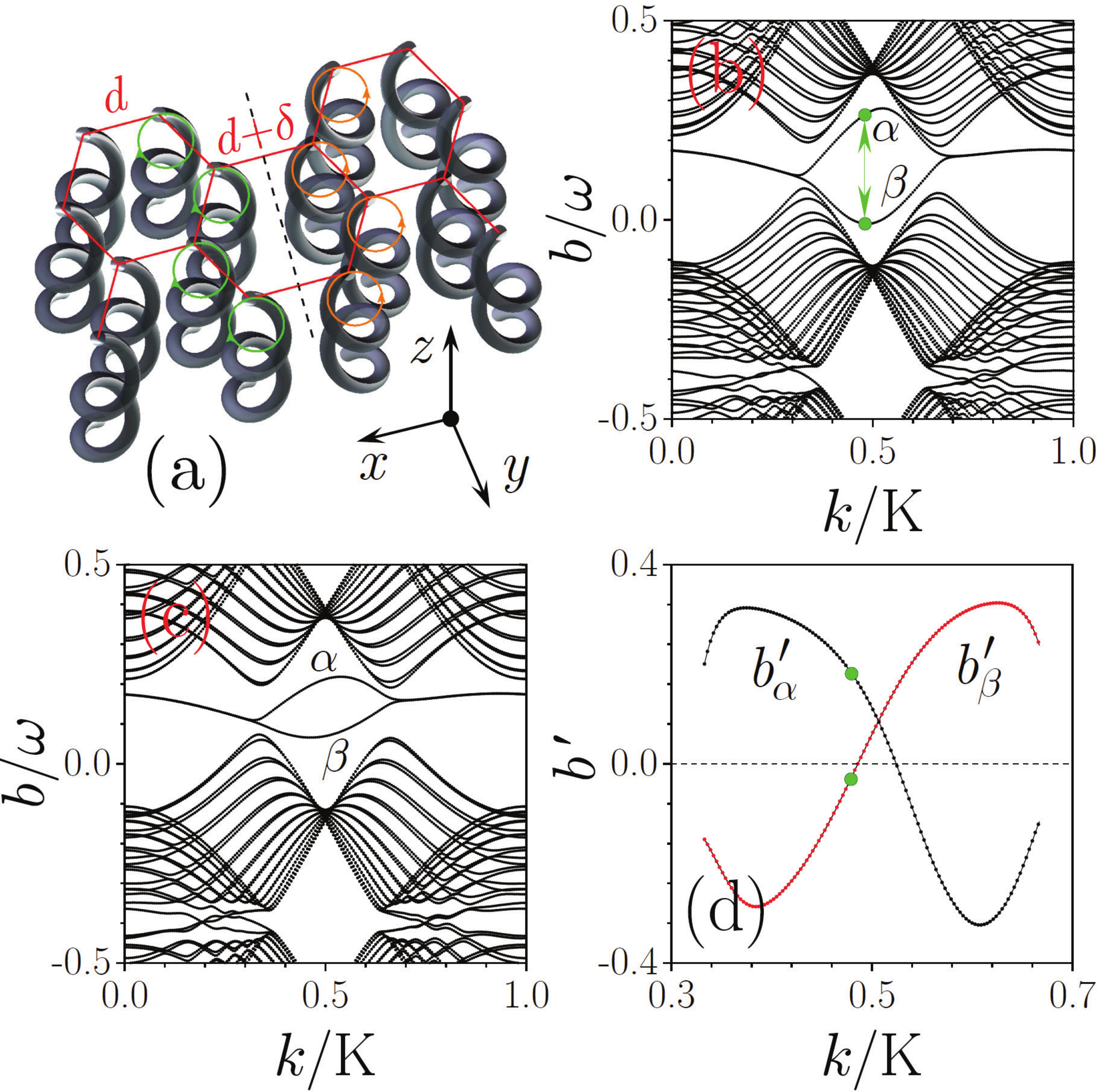}
\caption{(a) Schematic illustration of two waveguide arrays with opposite helicities shown by circles with arrows. Propagation constants $b$ of the eigenmodes versus $k/\textrm{K}$ for $Z=10$, $r_0=0.4$, {(b) $\delta=0.6$ and (c) $\delta=0.9$}. $\alpha$ and $\beta$ branches are associated with topological edge states at the zigzag-zigzag interface [dashed line in (a)]. Green arrow and dots in (b) indicate the states resonantly coupled by longitudinal modulation (see the text). (d) 
The $b'_{\alpha,\beta}$ for topological branches {at $\delta=0.6$.}}
\label{fig:one}
\end{figure}

In the paraxial approximation the propagation of a light beam along the $z$-direction in a medium with inhomogeneous refractive index creating optical potential $U(\br,z)$, is described by the NLS equation for the dimensionless field amplitude $\psi$:
\begin{equation} 
\label{NLS_dimensionless}
    i\partial_z\psi= -(1/2)\nabla^2\psi + [U(\br,z)+ U_d(\br,z)]\psi-|\psi|^2\psi.  
\end{equation}
Here $\br=x\hat{\bi}+y\hat{\bj}$, $\nabla=(\partial_x,\partial_y)$, $U_d(\br,z)$ is the additional weak modulation (discussed below),
and the last term describes the focusing Kerr nonlinearity of the medium.
We consider zigzag-zigzag interface of two honeycomb arrays with opposite waveguide rotation directions. Left and right arrays are located respectively at $x<0$ and $x>0$ [Fig.~\ref{fig:one}(a)]. The structure is periodic along the $y$- and $z$-axes with the periods $L$ and $Z$, respectively, i.e., $U(\br,z)=U(\br,z+Z)=U(\br+L\hat{\bj},z)$. The potentials created by the left, $U_-({\bm r},z)$, and right, $U_+({\bm r},z)$, arrays have the form
\begin{equation}
\label{Potential}
         U_\pm({\bm r},z)  = -p\sum_{m,n} e^{
         -\left|{\bm r}-{\bm r}_{nm}\mp \hat{\bi} (\delta+d)/2-r_0 {\bm s}(\pm z)\right|^4/h^4,
         }   
\end{equation}
  where $ {\bm s}(z)=\left(\sin(\omega z),\cos(\omega z)\right)$ describes helicity of the waveguides, ${\bm r}_{mn}$ are the coordinates of the knots of the honeycomb lattice identified by the integers $m$ and $n$, $h$ is the width of a single waveguide, $r_0$ is the helix radius, 
  $Z=2\pi/\omega$ is its period, $p$ is the depth of the potential, and $d$ is the distance between neighbour waveguides in each array (respectively $L=3^{1/2}d$). Additional spacing $\delta$ along the $x$-axis is introduced at the interface between the arrays [Fig.~\ref{fig:one}(a)] as a parameter controlling the linear spectrum  
  at the zigzag-zigzag interface at $x=0$ {[c.f. Figs.~\ref{fig:one}(b), (c)].}
  The total potential reads as $U({\bm r},z)=U_-({\bm r},z)+U_+({\bm r},z)$.

In the dimensionless units used here $\br$ is scaled to a characteristic width $w$, while $z$ is scaled to the diffraction length $2\pi w^2/\lambda$, where $\lambda$ is the wavelength. The potential depth is given by $p=\textrm{max}(\delta n/n)(2\pi w/\lambda)^2$. For an experimental implementation with fs-laser written waveguide arrays in fused silica (nonlinear coefficient $n_2\sim 1.4\cdot 10^{-20}~\textrm{m}^2/\textrm{W}$), we select $d=1.6$, $h = 0.4$, and $p=8.9$ corresponding to $16~\mu\textrm{m}$ separation between neighbouring waveguides of width $4~\mu\textrm{m}$ (we use $w=10~\mu \textrm{m}$),  refractive index contrast $6.5\cdot 10^{-4}$ at $\lambda=800~\textrm{nm}$, and propagation length of $z=1$ corresponding to $1.14~\textrm{mm}$. Helix radius $r_0=0.4$ ($4~\mu \textrm{m}$) and longitudinal period $Z=10$ ($11.39~\textrm{mm}$) guarantee relatively low radiative losses. {Weak losses $\sim 0.1$ dB/cm in waveguides inscribed in such transparent media lead only to slow broadening of solitons, whose width self-adjusts to gradually decreasing peak amplitude. Below we omit the effect of losses upon discussion of the existence of Bragg solitons.}

The linear Hamiltonian of unperturbed helical arrays, $H_0(\br,z)\equiv -(1/2)\nabla^2 + U(\br,z)$, is characterized by a Floquet spectrum $b_{\gamma k}\in [-\omega/2,+\omega/2]$, where $\omega=2\pi/Z$, $k$ is the Bloch momentum along the $y$-axis in the reduced Brillouin zone of width $\textrm{K}=2\pi/L$, and $\gamma$ enumerates allowed bands and edge states (if any). Each Floquet state $\psi_{\gamma k}(\br,z)=\phi_{\gamma k}(\br,z)e^{ib_{\gamma k}z}$ is associated with (quasi-) propagation constant $b_{\gamma k}$, where  $\phi_{\gamma k}=u_{\gamma k}(\br,z)e^{iky}$, and $u_{\gamma k}$ is the $L$- and $Z$-periodic function: $u_{\gamma k}(\br,z)=u_{\gamma k}(\br,z+Z)=u_{\gamma k}(\br+L\hat{\bj},z)$. Figures ~\ref{fig:one}(b,c) show representative Floquet spectra of the described arrays. The spectrum features two branches of the edge states, denoted by $\alpha$ and $\beta$ at $\textrm{K}/3<k<\textrm{2K/3}$ {(these values are associated with projections on the $k_x=0$ plane of Dirac cones at \textbf{K} and \textbf{K'} corners of full Brillouin zone for bulk honeycomb arrays)}. These states emerge only for nonzero separation between arrays $\delta>0$. {The branches are well separated for $\delta\sim d$, but they gradually approach each other when separation between the arrays increases [cf. Figs. \ref{fig:one}(b) and \ref{fig:one}(c)]}. The branches $\alpha$ and $\beta$ are characterized by different 
derivatives $b'_{\alpha,\beta}=\partial b_{\alpha,\beta}/\partial k$ shown in Fig. ~\ref{fig:one}(d).

Any two modes belonging to these branches can be resonantly coupled by small periodic modulation $U_d$ of the main structure if the modulation ensures the conservation of propagation constant and Bloch momentum. Small modulation $U_d$ can be transverse (along $y-$ axis) allowing for matching Bloch momenta, or longitudinal (along $z-$axis) allowing matching propagation constants. Here we focus on the latter type of modulation, because transverse one proved to be inefficient for the purpose of generation of Bragg solitons in helical arrays. In Eq.~(\ref{NLS_dimensionless}) weak longitudinal modulation is described by the additional potential $U_d(\br,z)=\mu \textrm{sin}(\Omega z)\mathrm{sign}(x)U(\br,z)$, where $\mu\ll1$ is a small parameter and $\Omega$ is the frequency of slow modulation, while $\textrm{sign}(x)$ ensures the most efficient coupling of the edge states having different $x$-symmetries. Such modulation couples two modes characterized by the same Bloch momentum $k$, e.g. modes  $\psi_{\alpha k}$ and $\psi_{\beta k}$ satisfying the exact matching condition $b_{\alpha,k}-b_{\beta,k}=\Omega$ [see the circles in Fig.~\ref{fig:one}(b); below we use that $b_{\alpha,k}>b_{\beta,k}$]. Such Rabi coupling is a resonant process (see e.g, \cite{valley1}, where this effect was reported for valley-Hall states).  

Next we use the  multiple-scale expansion, modified to take into account weak periodic modulations along the evolution direction {\cite{our-ACSphot}}. We search for a solution of (\ref{NLS_dimensionless}) in the form
 \begin{eqnarray}
\label{psi_expan_time}
\psi=  [a_{\alpha} (y,z)\psi_{\alpha k}
+a_{\beta} (y,z)\psi_{\beta k}] 
\nonumber \\
+[c_{\beta}^{(\alpha)} (\br,z)\phi_{\beta k}e^{ib_{\alpha k} z}+c_{\alpha}^{(\beta)} (\br,z)\phi_{\alpha k}e^{ib_{\beta k} z}]
+  \tpsi(\br,z). 
\end{eqnarray}
Here  $a_{\alpha,\beta} (y,z)$ are the slowly varying amplitudes of the respective edge states, which additionally must be small enough to ensure the scaling relations
$|\partial_z a|\sim |\partial_y a| \sim |a|^3\sim\mu |a| \ll \Omega |a|$. The terms proportional to $c_{\alpha}^{(\beta)}$ and $c_{\beta}^{(\alpha)}$ (with $|c|\ll |a|$) describe excitations accompanying soliton evolution (they can be termed companion modes~\cite{Sterke}). 
Unlike in the case of vector gap solitons in stationary periodic medium (see e.g.~\cite{VK}), now the companion terms
acquire the same propagation constants as main soliton components, and can also be resonantly coupled by the periodic longitudinal modulation. Even if such modes do not exist at the input of the array - they are excited during propagation (see (\ref{c}), below). The $\tpsi(\br,z)$ describes the rest of the higher order terms,
$|\tpsi|\sim |c|\ll |a|$. 

In what follows, we drop details of lengthy calculations (they are analogous to those performed in \cite{optsol4}) and only present the most relevant final expressions. For the envelopes we obtain
\begin{eqnarray}
\label{coupled_z}
\begin{array}{l}
 \displaystyle{
 i\frac{\partial a_{\alpha}}{\partial z}- i b_\alpha^\prime\frac{\partial a_{\alpha}}{\partial y}  +\kappa a_{\beta} +\left(\chi_\alpha |a_\alpha|^2+2\chi  |a_\beta|^2\right)a_\alpha =0 
 } 
 \\[0.2cm]
 \displaystyle{
i\frac{\partial a_{\beta}}{\partial z}- i b_\beta^\prime \frac{\partial a_{\beta}}{\partial y}  +\kappa^* a_{\alpha} +\left(\chi_\beta |a_\beta|^2+2\chi  |a_\alpha|^2\right)a_\beta =0   
}
\end{array}
\end{eqnarray}
Here $\kappa=\langle (u_{\alpha k},U_du_{\beta k})\rangle_Z$ is the coupling coefficient, $\chi_{\gamma}=\langle (|u_{\gamma k}|^2, |u_{\gamma k}|^2)\rangle_Z$ with $\gamma=\alpha,\beta$ are the self-phase-modulation coefficients, $\chi=\langle (|u_{\alpha k}|^2, |u_{\beta k}|^2)\rangle_Z$ is the cross-phase modulation, the $z-$averaging is defined by $\langle f\rangle_Z=(1/Z)\int_{0}^{Z}f(z)dz$,  the  inner product is  $(f_1,f_2)=\int_S f_1^*f_2d\br$, where $S$ is the transverse area of the array, and the normalization condition for the eigenmodes $(\phi_{\gamma' k'},\phi_{\gamma k})  =\delta_{\gamma\gamma'}\delta_{kk'}$ is used.  The effective dispersion is related to the Floquet modes $b_\gamma^\prime=\langle \left(\phi_{\gamma},i\partial_ {y}\phi_{\gamma}\right)\rangle_Z$.

System (\ref{coupled_z}) admits Bragg soliton solutions~\cite{Bragg,Aceves,optsol4}:
\begin{equation}
 \label{soliton}
 \left(\!
 \begin{array}{c}
      a_\alpha  \\
      a_\beta
 \end{array}
 \!\right)= \frac{\sqrt{2|\kappa|}\tau e^{i\varphi+i\eta }\sin\sigma}{\sqrt{\tau^8\chi_\beta+4\tau^4\chi+\chi_\alpha}} 
 \left(\!
 \begin{array}{c}
 \mbox{sech}(\xi+i\sigma/2)e^{i\phi/2}
 \\
 \tau^2\mbox{sech}(\xi-i\sigma/2)e^{-i\phi/2}
 \end{array}
 \!\right)
 \end{equation}
 where $\zeta_{\alpha,\beta}=y+b_{\alpha,\beta}^\prime z$,
 \begin{eqnarray*}
 	\xi=\frac{|\kappa| \left(\tau^4\zeta_\beta+\zeta_\alpha\right)\sin\sigma}{\tau^2 (b_\beta^\prime-b_\alpha^\prime)} , 
 	\quad
 	\varphi=\frac{|\kappa| \left(\tau^4\zeta_\beta-\zeta_\alpha\right)\cos\sigma }{\tau^2(b_\beta^\prime-b_\alpha^\prime)} 
 	\\
 	\eta=\frac{2(\tau^8\chi_\beta-\chi_\alpha) }{\chi_\alpha+4 \chi\tau^4+\chi_{\beta}\tau^8}\arctan\left( \frac{1-\cos\sigma}{\sin\sigma}\tanh\xi\right)
 \end{eqnarray*}
and $\sigma\in(0,\pi/2)$ and $\tau\in(-1,1)$ are the free parameters. 
The soliton (\ref{soliton}) moves with velocity of the envelope: $v_\textrm{sol}=-{(b_\alpha^\prime+\tau^4 b_\beta^\prime)}/{(1+\tau^4)}$, which is {\em  different} from the individual effective group velocities of the coupled modes given by $-b_{\alpha\beta}^\prime$.
 
 The amplitudes of the companion modes, which are zero at $z=0$, 
 are given by
\begin{equation}
\label{c}
\begin{split}
c_{\beta}^{(\alpha)}=&iv_{\beta\alpha}\frac{e^{i(b_\beta-b_\alpha)z}-1}{b_\beta-b_\alpha}\frac{\partial a_{\alpha}}{\partial y},
\\
c_{\alpha}^{(\beta)}=&iv_{\alpha\beta}\frac{e^{i(b_\alpha-b_\beta)z}-1}{b_\alpha-b_\beta}\frac{\partial a_{\beta}}{\partial y},
\end{split}
\end{equation}
where $v_{\alpha\beta}=-\langle(\phi_{\alpha k},i\partial_{y}\phi_{\beta k} )\rangle_Z$. Thus, the companion modes are resonantly excited during the evolution of the soliton (\ref{soliton}).  
\begin{figure}[t!]
\centering
\includegraphics[width=1\linewidth]{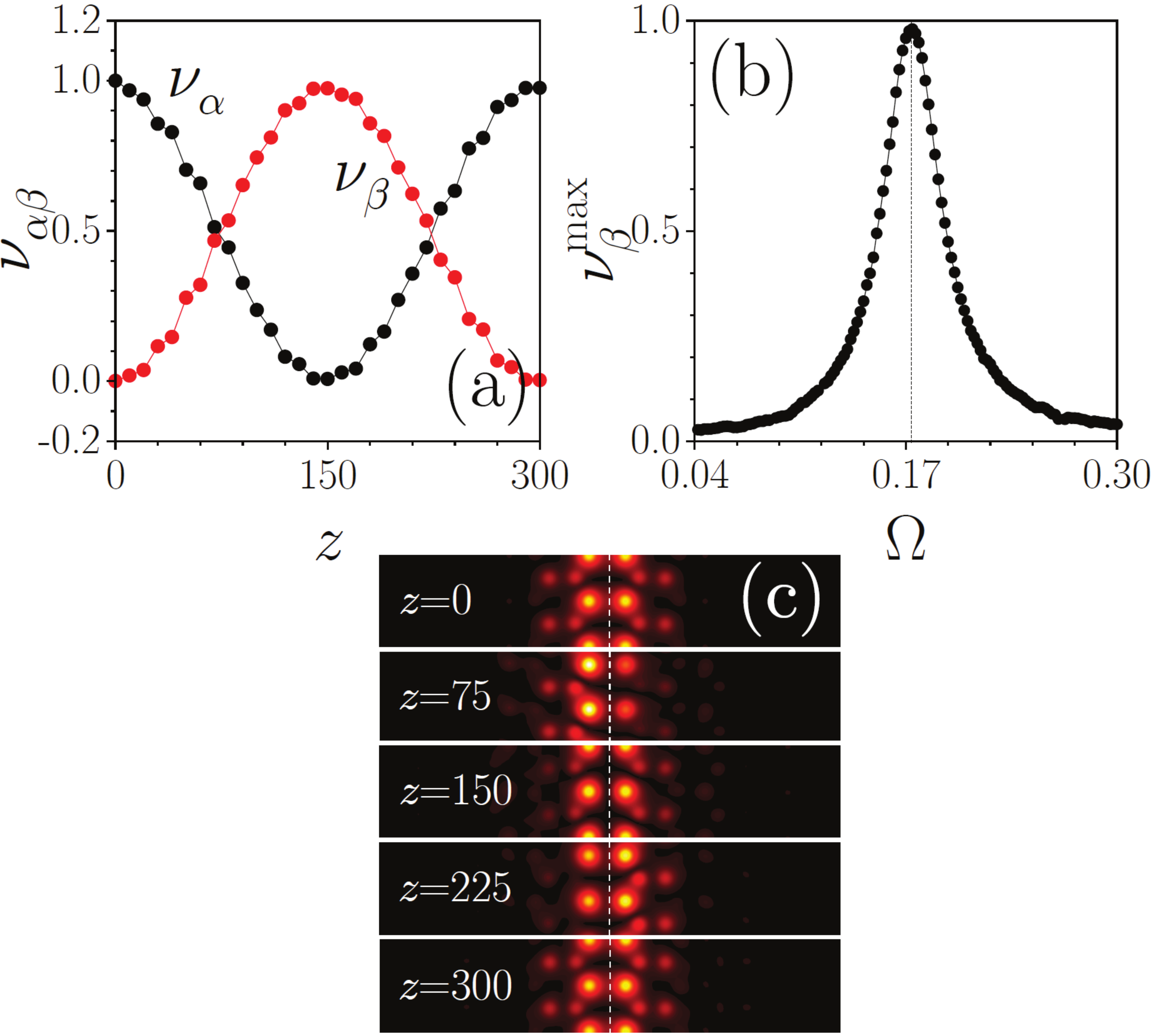}
\caption{(a) Modal weights illustrating resonant transition between edge states $\alpha,\beta$ at $\Omega=0.172$. (b) Resonance curve showing maximal weight $\psi^\textrm{max}_\beta$ of the edge state $\beta$ as a function of modulation frequency $\Omega$. (c) $|\psi|$ distributions at different distances illustrating mode transformation for parameters of panel (a). In all cases $k=0.48\textrm{K}$, $\mu=0.005$.}
\label{fig:two}
\end{figure}

\begin{figure}[t!]
\centering
\includegraphics[width=1\linewidth]{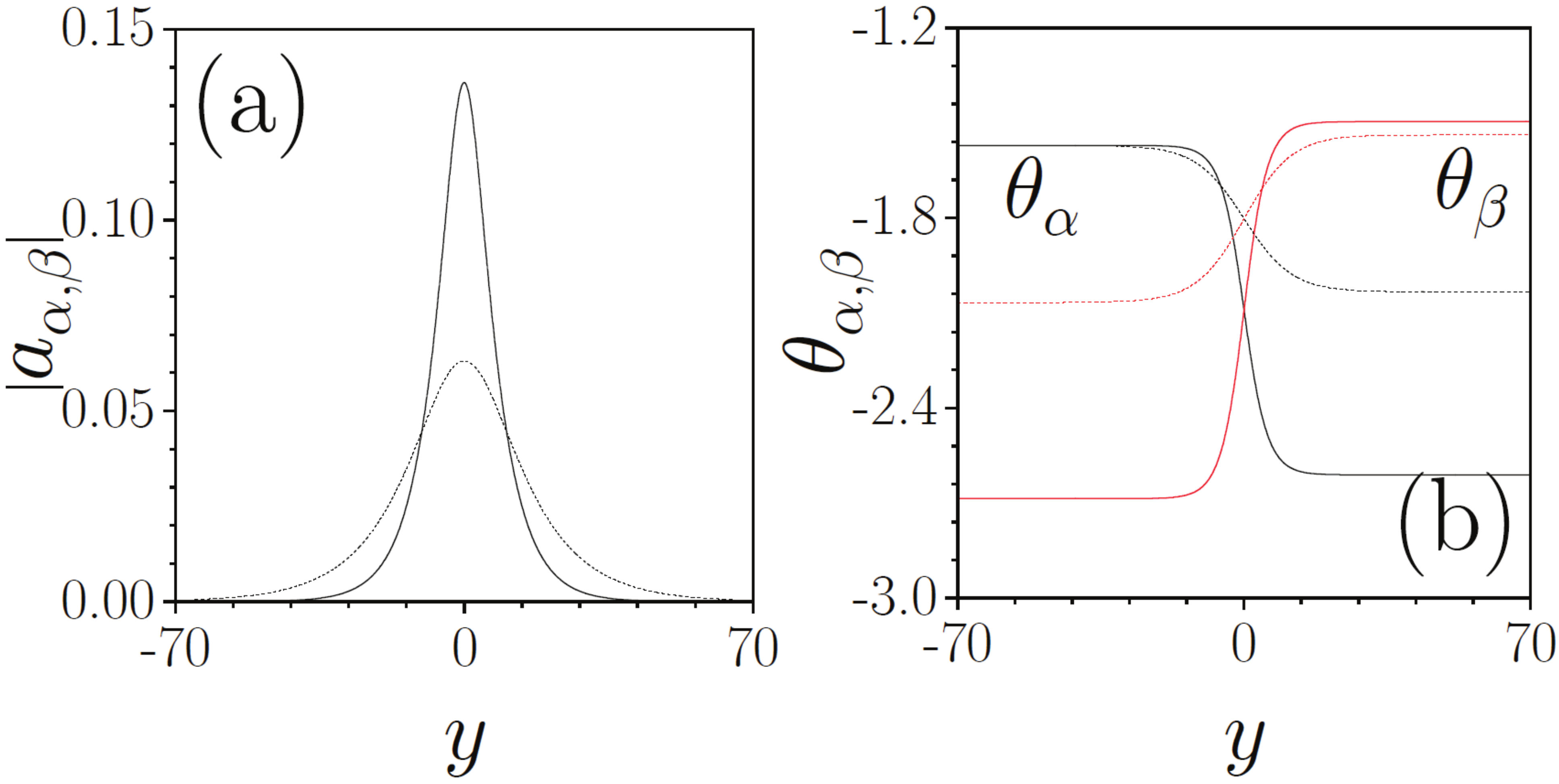}
\caption{Envelope modulus $|a_\alpha|=|a_\beta|$ (a) and phase $\theta_{\alpha,\beta}$ (b) distributions in Bragg solitons corresponding to $\chi_\alpha = 0.334$, $\chi_\beta = 0.408$, $\chi = 0.365$, $b_\alpha^\prime = 0.181$, $b_\beta^\prime = - 0.0255$, $\tau = 1$ and $\sigma = 0.49$ (dashed lines) and $\sigma = 0.88$ (solid lines) at $|\kappa|=0.0182$.}
\label{fig:three}
\end{figure}

\begin{figure*}[t]
	\centering
	\includegraphics[width=\textwidth]{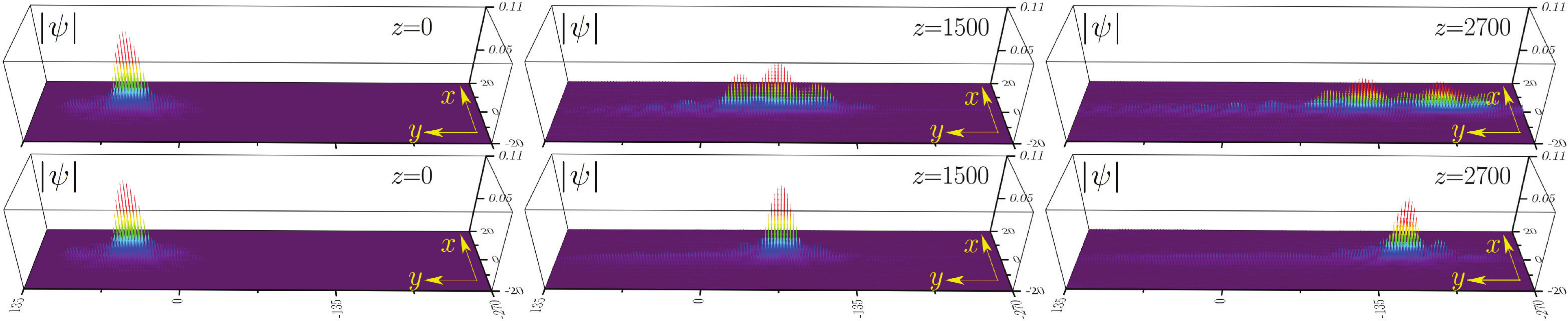}
	\caption{Propagation of the topological Bragg quasi-soliton constructed for $\chi_\alpha = 0.334$, $\chi_\beta = 0.408$, $\chi = 0.365$, $|\kappa|=0.0182$, $\sigma = 0.49$, $\tau = 1$, and moving with velocity $-(b'_\alpha+b'_\beta)/2$ 
		where $b_\alpha^\prime = 0.181$, $b_\beta^\prime = - 0.0255$, in the linear (top row) and nonlinear (bottom row) media.}
	\label{fig:four}
\end{figure*}

\begin{figure}[t!]
	\centering
	\includegraphics[width=\linewidth]{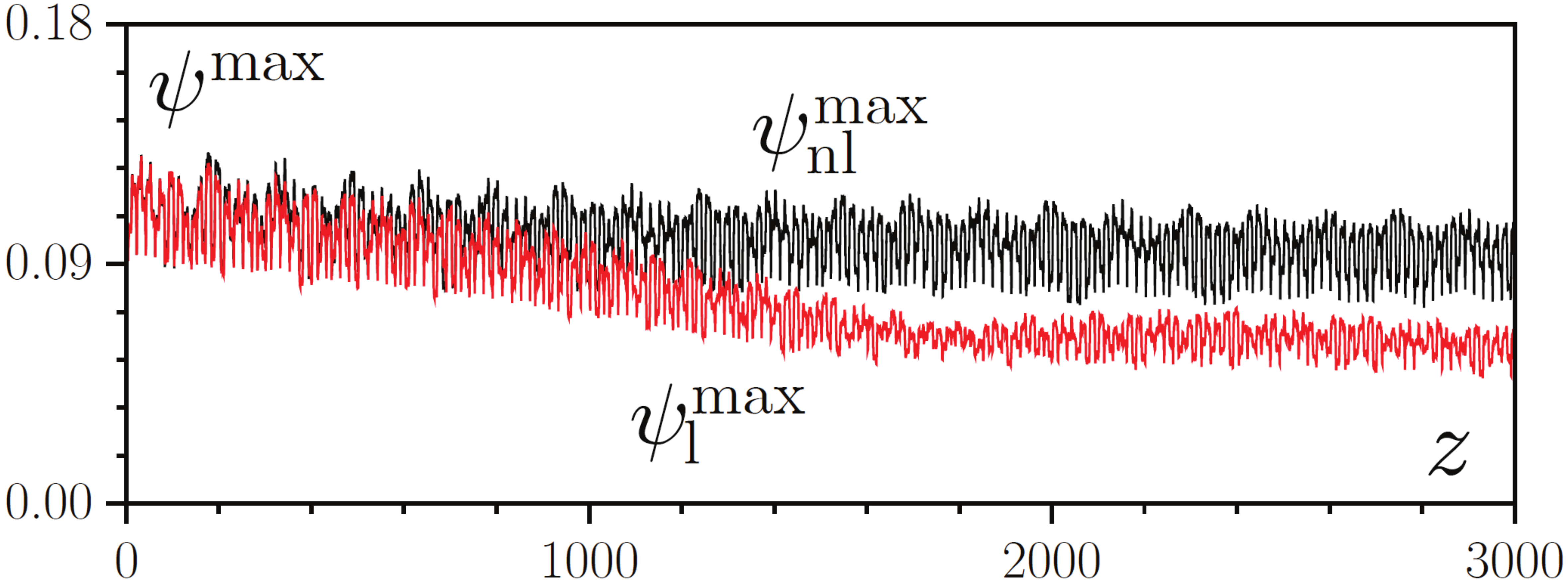}
	\caption{Peak amplitudes versus $z$ for linear and nonlinear dynamics shown in Fig.~\ref{fig:four}.}
	\label{fig:five}
\end{figure}

To illustrate resonant coupling between topological edge states, induced by the weak longitudinal modulation $U_d(\br,z)$, in Fig.~\ref{fig:two}(a) we show Rabi oscillations of the intensities $\nu_{\alpha,\beta}=|a_{\alpha,\beta}|^2$, where the amplitudes $a_\gamma$ were computed as projections $(\psi_{\gamma k},\psi(\br,z))$ upon direct numerical simulations of continuous Eq.~(\ref{NLS_dimensionless}). The intensities of the modes computed at discrete steps equal to the helix period $Z$ are shown by the dots. As one can see from the figure, the process is periodic with a period approximately equal to analytically predicted period $\pi/|\kappa|$. Even weak modulation with $\mu=0.005$ leads to practically complete transition between two edge states in resonance $\Omega=b_{\alpha k}-b_{\beta k}\approx 0.172$. {The states $\alpha$ and $\beta$ have different $x$-symmetries, thus their selective excitation is possible by using beams with proper symmetries whose momentum $k$ is matched to that of the edge state.} Deviation from resonance modulation frequency results in decrease of the coupling efficiency. The dependence of the maximal intensity $\nu_\beta^\textrm{max}$ of the $\beta$ state on the modulation frequency $\Omega$ is shown in Fig.~\ref{fig:two}(b). One can see that maximal intensity $\nu_\beta^\textrm{max}$ rapidly decreases away of the resonance at $\Omega=0.172$. The evolution of the field distribution corresponding to panel (a) is illustrated in Fig. \ref{fig:two}(c) at the distances $z=0, \ \pi/4|\kappa|, \ \pi/2|\kappa|, \ 3\pi/4|\kappa|, \  \pi/|\kappa|$. Notice change of the symmetry of the wave at $z=0$ (nearly pure $\alpha$ state) and $z\approx\pi/2|\kappa|$ (nearly pure $\beta$ state).

Next, we focus on Bragg quasi-solitons. The representative envelopes of such solitons given by Eq. (\ref{soliton}) are shown in Fig.~\ref{fig:three} for $k=0.48\textrm{K}$ and $|\kappa|=0.0182$ (corresponding to $\mu=0.01$). For fixed $\tau=1$ corresponding to the soliton velocity $v_\textrm{sol}=-(b_\alpha^\prime+ b_\beta^\prime)/2$ the amplitude of Bragg soliton increases with $\sigma$ [Fig.~\ref{fig:three}(a)], and so does the step in soliton phase $\theta_{\alpha,\beta}$ that most rapidly varies in the region with large intensity [Fig.~\ref{fig:three}(b)]. Thus, in contrast to conventional Schr\"odinger quasi-solitons, Bragg ones are always chirped. {The solitons constructed using envelopes from (\ref{soliton}) were excited within a wide range of input peak amplitudes from 0.05 to approximately 0.30}. Representative propagation dynamics of a Bragg soliton, which at the input is given by a superposition of the edge states: $\psi_{in}=  a_{\alpha} (\br,z_{in})\psi_{\alpha k}(\br,z_{in})+a_{\beta} (\br,z_{in})\psi_{\beta k}(\br, z_{in})$, c.f. (\ref{psi_expan_time}), where $z_{in}<0$, is illustrated in Fig.~\ref{fig:four}. The top (bottom) row of this figure shows evolution without (with) nonlinearity. {The evolution is shown for the interval $z\in[0,3000]$, after a short transient period, $z\in[z_{in},0]$, where $z_{in}=-100$, during which the input beam $\psi_{in}$ looses an appreciable amount of its energy. These losses are due to the fact that the initial state $\psi_{in}$ predicted by the theory provides a $z$-averaged approximation for the exact solution, which always performs small oscillations due to its Floquet nature 
as it is illustrated 
in Fig.~\ref{fig:five}. Thus, during a transient period an initial approximation transforms into exact oscillating solution.} The formation of a Bragg soliton that moves along the edge over considerable distance, experiencing only moderate amplitude oscillations is obvious from the bottom row of Fig.~\ref{fig:four}. In contrast, the same input dramatically disperses in the linear medium, as shown in the top row. The difference in linear and nonlinear dynamics is also illustrated in Fig.~\ref{fig:five}, where we compare evolution of peak amplitudes $\psi^{\rm max}$ in the linear and nonlinear cases. {While further growth of input amplitude increases contrast with linear evolution, it may drive the excitation into the band causing unwanted coupling with bulk modes}.

For the Bloch momentum $k=0.48\textrm{K}$ the branches $\alpha$ and $\beta$ feature different signs of the second-order dispersion $b''_{\alpha,\beta}=\partial^2 b_{\alpha,\beta}/\partial k^2$.  For equal signs of the effective nonlinearities $\chi_{\alpha,\beta}$ and $\chi$, this excludes the possibility of the formation of  conventional  bright vector Schr\"odinger solitons. In conjunction with specific soliton velocity coinciding with its theoretical prediction and with the scaling relations between the amplitude and width of the obtained solitons, this confirms Bragg nature of the excited nonlinear states. Notice also, that even though Bragg solitons survive over considerable propagation distances, they are quasi-solitons, i.e. their peak amplitude very slowly decreases with distance [Fig.~\ref{fig:five}] due to small radiation, visible in the bottom row of [Fig.~\ref{fig:four}]. This phenomenon, is due to the companion modes: although those modes are much smaller than $\psi_{in}$, they are localized around the soliton. Having the same propagation constants $b_{\alpha\beta}$, companion modes introduce higher dispersion, thus breaking balance with the nonlinearity.

Summarising, we have predicted the existence of new type of Bragg edge quasi-solitons in Floquet topological insulators and developed their theory. {Our results suggest that Bragg solitons may form in other systems supporting more than one edge state per interface, such as gyromagnetic photonic crystals and atomic systems with spin-orbit coupling.}

\medskip

\noindent\textbf{Funding Information.} 
DFG (grant SZ 276/19-1);  RFBR (grant 18-502-12080); Portuguese Foundation for Science and Technology (FCT) under Contract no. UIDB/00618/2020. 

\medskip

\noindent\textbf{Disclosures.} The authors declare no conflicts of interest.


\end{document}